\documentclass{aa}

\usepackage{graphicx}
\usepackage{txfonts}                 %  Postscript Fonts
\newcommand{\kms}{km\,s$^{-1}$}
\newcommand{\Halpha}{H$\alpha$}

\usepackage{txfonts}
\begin{document}

\titlerunning{II Peg with PEPSI}
\authorrunning{Strassmeier, Carroll \& Ilyin}

\title{Warm and cool starspots with opposite polarities\thanks{Based on data acquired with
    PEPSI using the Large Binocular Telescope (LBT). The LBT is an
    international collaboration among institutions in the United States,
    Italy, and Germany.}}

   \subtitle{A high-resolution Zeeman-Doppler-Imaging study of II Pegasi with PEPSI}

   \author{K.~G.~Strassmeier\inst{1}
        \and T.~A.~Carroll\inst{1}
        \and I.~V.~Ilyin\inst{1}
   }

   \institute{Leibniz-Institut f\"ur Astrophysik Potsdam (AIP),
     An der Sternwarte 16, 14482 Potsdam, Germany\\
              \email{kstrassmeier@aip.de}
}

   \date{Received xxx x, 2018; accepted xxx x, 2019}

  \abstract
  % context heading (optional)
  % {} leave it empty if necessary
   {}
  % aims heading (mandatory)
   {We present a temperature and a magnetic-field surface map of the K2 subgiant of the active binary II~Peg. Employed are high resolution Stokes IV spectra obtained with the new Potsdam Echelle Polarimetric and Spectroscopic Instrument (PEPSI) at the Large Binocular Telescope (LBT).}
  % methods heading (mandatory)
   {Fourteen average line profiles are inverted using our $i$Map code. We have employed an iterative regularization scheme without the need of a penalty function and incorporate a physical 3D description of the surface field vector. The spectral resolution of our data is 130\,000 which converts to 20 resolution elements across the disk of II~Peg. }
   % results heading (mandatory)
   {Our main result is that the temperature features on II~Peg closely correlate with its magnetic field topology. We find a warm spot ($350$\,K warmer with respect to the effective temperature) of positive polarity and radial field density of 1.1\,kG coexisting with a cool spot ($780$\,K cooler) of negative polarity of 2~kG. Several other cool features are reconstructed containing both polarities and with (radial) field densities of up to 2\,kG. The largest cool spot is reconstructed with a temperature contrast of $550$\,K, an area of almost 10\% of the visible hemisphere, and with a multipolar magnetic morphology. A meridional and an azimuthal component of the field of up to $\pm$500\,G is detected in two surface regions between spots with strong radial fields but different polarities. A force-free magnetic-field extrapolation suggests that the different polarities of cool spots and the positive polarity of warm spots are physically related through a system of coronal loops of typical height of $\approx$2\,R$_\star$. While the \Halpha\ line core and its red-side wing exhibit variations throughout all rotational phases, a major increase of blue-shifted \Halpha\ emission was seen for the phases when the warm spot is approaching the stellar central meridian indicating high-velocity mass motion within its loop. }
  % conclusions heading (optional), leave it empty if necessary
   {Active stars such as II~Peg can show coexisting cool and warm spots on the surface that we interpret resulting from two different formation mechanisms. We explain the warm spots due to photospheric heating by a shock front from a siphon-type flow between regions of different polarities while the majority of the cool spots is likely formed due to the expected convective suppression like on the Sun. }

   \maketitle
%
%-------------------------------------------------------------------

\section{Introduction}

\object{II~Peg} (HD~224085) is among the most active and spotted RS~CVn binaries known and its early-K subgiant component one of the magnetically most active cool stars in the sky. Already 40 years ago, it had been center of many studies across the entire electromagnetic spectrum, focussing on its extremely spotted photosphere (e.g., Rucinski \cite{ruc}, Vogt \cite{vogt79}), its chromosphere (e.g., Byrne et al. \cite{byrne}), corona (e.g., Schrijver et al. \cite{schr}, Mutel \& Lestrade \cite{mut:les}), large flares (e.g., Doyle et al. \cite{doyle}) or its magnetism in general (e.g., Vogt \cite{vogt80}). It was among the first targets for which systematically Doppler imaging (DI) was  applied. Temperature and/or brightness maps were presented by Xiang et al. (\cite{xi}) for 2004, Hackman et al. (\cite{hack12}) for 2004-2010, by Lindborg et al. (\cite{lind11}) for 1994-2002, by Carroll et al. (\cite{carr07}) for 2004, by Gu et al. (\cite{gu}) for 1999-2001, by Berdyugina et al. (\cite{berd99}) for 1997-1998, by Weber (\cite{web04}) for 1996-1997, by Berdyugina et al. (\cite{berd98}) for 1992-1996, by Zboril (\cite{zbor03}) for 1993, and by Hatzes (\cite{hatz95}) for 1992-1994.

Direct magnetic field measurements from Zeeman splitting is not possible for II~Peg due to its rotational line broadening of $\approx$22~\kms. The first detection of its magnetic field was achieved with a differential technique by Vogt (\cite{vogt80}) which consisted of a pair of consecutive spectra, the first spectrum right-circularly polarized and the second left-circularly polarized (essentially Stokes V). In this way, a longitudinal magnetic field of strength $-515$~G was detected, but not regarded significant. However, Donati et al. (\cite{don92}) confirmed the detection later from independent Stokes V data but with an effective magnetic field density never exceeding 70~G.

The technique of Zeeman-Doppler imaging (ZDI) was introduced three decades ago by Semel (\cite{semel}) and had become an indispensable technique for stellar magnetic field studies as noticed in many review papers (e.g., Strassmeier \cite{spots}, Donati \& Landstreet \cite{don:lan}, Reiners \cite{reiners}). Among its main advantages is that it resolves the polarity of magnetic features. However, it took almost two decades that the first ZDI of II~Peg was constructed (Carroll et al. \cite{carr07} from SOFIN Stokes IV data). Their ZDI map revealed local radial-field densities of up to $\pm$600~G and was only possible with a multi-line principal component reconstruction because of the comparably weak polarimetric signatures. Still, it is noted that the surface field exhibited a strong imbalance in the magnetic polarity and we believe that this imbalance of magnetic flux results, at least partially, from the inability of the circular polarization signal to capture the whole magnetic flux of the star. A more extensive Stokes~IV study followed by Kochukhov et al.~(\cite{koc:man}) and revealed a complex and evolving surface magnetic-field morphology on II~Peg.

Linear polarization (LP) in spectral lines of the Sun is typically up to ten times weaker than circular polarization (CP) (e.g., Stenflo \cite{stenflo89}) and hardly accessible with reasonable signal-to-noise ratio (S/N) in stars. However, Ros\'en et al. (\cite{rosen13}) found strong and variable LP in II~Peg. Ros\'en et al. (\cite{rosen15}) also noted that the LP signal ``cannot be readily seen in individual spectral lines at the S/N of our observations'' and thus made the application of least-square deconvolution mandatory if to be included in the ZDI inversion. Nevertheless, this spurred the first ZDI of a cool star using all four Stokes parameters (Ros\'en et al. \cite{rosen15}). It also led to a first, and so far only, comparison of a ZDI map generated from Stokes~IV with a map from Stokes IQUV. The field strength of surface features was doubled or even quadrupled with respect to the Stokes IV map when LP was taken into account, which the authors accounted to an increase in the overall complexity of the magnetic morphology. Same holds for the total magnetic energy because the reconstructed field got more complex when IQUV was used.

%------------------------F1
   \begin{figure*}
   \centering
   \includegraphics[angle=0,width=\textwidth, clip]{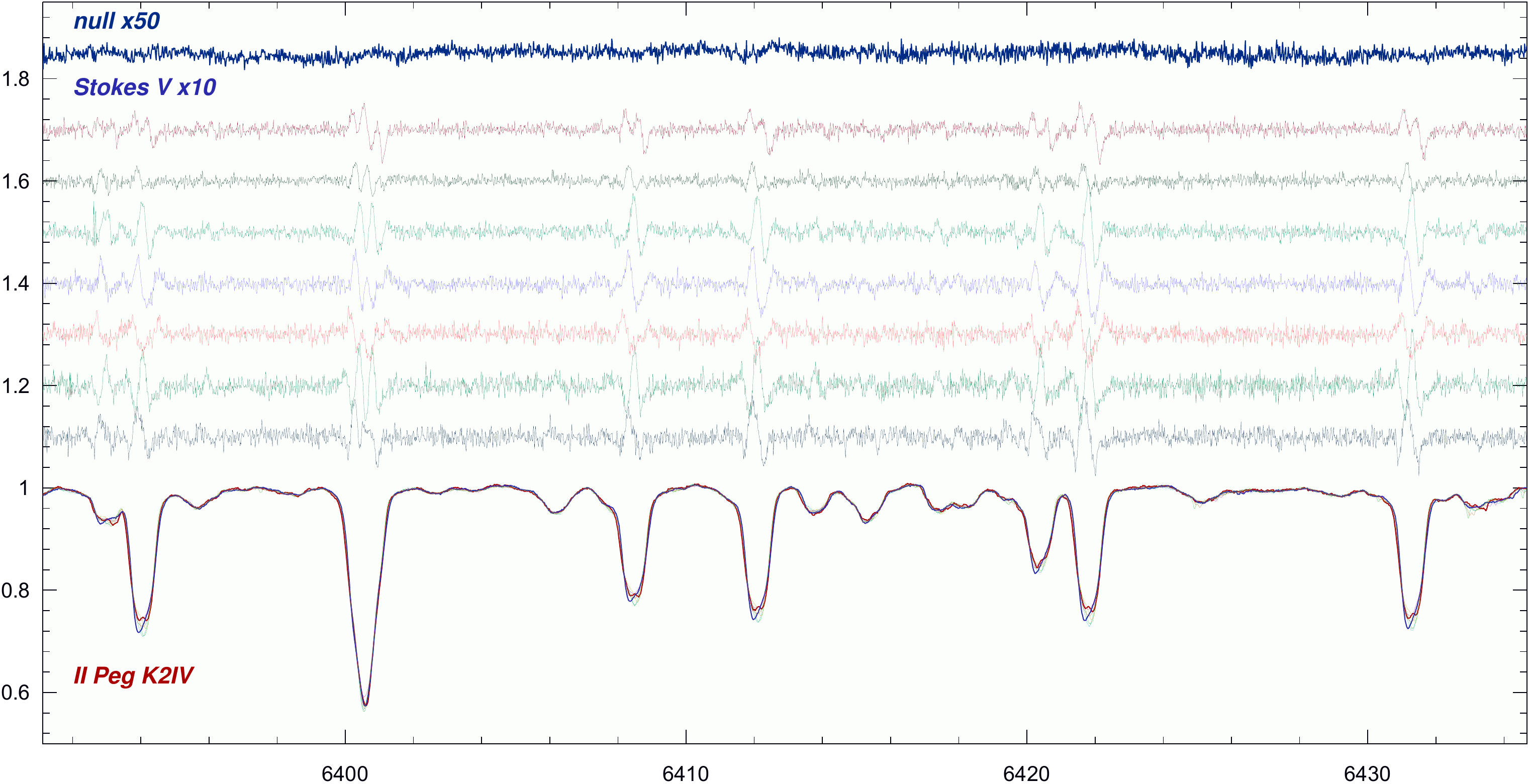}
   \caption{Spectra for an example wavelength region around 6400\,\AA . The bottom spectra are seven overplotted Stokes~I spectra for one spectrum per night (out of the 14 rotational phases of the full time series). Note that the rotation period of II~Peg is 6.7~d. On top of it are the same seven Stokes~V spectra for the seven consecutive nights but shifted in relative intensity by 0.1 and enhanced in scale by a factor of ten with respect to Stokes~I. Time increases from bottom to top. The very top spectrum is the null spectrum enhanced by a factor 50 with respect to Stokes~I.}
         \label{F1}
   \end{figure*}

In this paper, we present a Zeeman Doppler image from data taken during the PEPSI-POL commissioning with the 2$\times$8.4\,m LBT. Although only Stokes I and V were available at that time in the commissioning, the spectral resolution is now 130\,000 and the LBT enabled a peak S/N of 1280 per pixel, that is, both quantities twice as high as for previous data. We found that the complexity of II~Peg's magnetic field appears increased from Stokes~IV alone with respect to our earlier lower-resolution map from SOFIN Stokes-IV data, which is in line with the findings based on the full Stokes vector by Ros\'en et al. (\cite{rosen15}). Our new observations are described in Sect.~2. Our ZDI code $i$Map is described in Sect.~3 along with its assumptions, the input parameters, the creation of singular-value decomposition (SVD) line profiles, and the resulting Doppler and Zeeman-Doppler images. Section~4 discusses the results and Sect.~\ref{S5} concludes and summarizes our findings.

%--------------------------------------------------------------------
\section{Observations}\label{S2}

The polarimetric observations in this paper were obtained with PEPSI at the 2$\times$8.4\,m LBT in Arizona. We employed both polarimeters in the LBT's two symmetric straight-through Gregorian foci. Two pairs of octagonal 200$\mu$m fibers per polarimeter feed the ordinary and extraordinary polarized beams via a five-slice image slicer per fiber into the spectrograph. It produces four spectra per \'echelle order with a 4.2-pix spectral resolution of $R$=130\,000 that are recorded in a single exposure. The two polarimeters are identical in design and construction but are separately calibrated. Both are based on a classical dual-beam design with a modified Foster prism as linear polarizer with two orthogonally polarized beams exiting in parallel. The Foster prism, atmospheric dispersion corrector (ADC), two fiber heads, and two fiber viewing cameras are rotating as a single unit with respect to the parallactic axis on the sky. The red-optimized polymethylmethacrylat (PMMA) quarter-wave retarder is inserted into the optical beam in front of the Foster prism for the CP measurements and retracted off for the LP measurements. This design allows us to avoid any cross-talk between CP and LP which would be introduced in case of a half-wave retarder (Ilyin \cite{ilya12}). The spectrograph and the polarimeters were described in detail by Strassmeier et al. (\cite{pepsi}, \cite{spie-austin}).

Observations of II~Peg commenced over seven consecutive nights as part of the instrument commissioning in October 2017. Fourteen spectra were taken simultaneously in two wavelength regions with cross disperser (CD) III covering 4800--5441~\AA\ and with CD~V covering 6278--7419~\AA . We used the two 8.4\,m LBT mirrors (dubbed SX and DX) independently, that is CP with SX and LP with DX simultaneously. Unfortunately, a problem with the DX ADC did not allow the scientific use of LP. Retarder angles of 45\degr\ and 135\degr\ were set for Stokes V, Foster prism position angles of 0\degr\ were set for Stokes Q and 45\degr\ for Stokes U with respect to the north. We note again that the retarder is removed from the beam during the LP observations. Exposure time per subintegration was 20\,min (15\,min and 25\,min for some exceptions). Average S/N per pixel was around 1000 in CD~V and 550 in CD~III for a pair of exposures. The Stokes~V polarization signatures in the spectra of II~Peg are easily recognized by eye. Figure~\ref{F1} is a plot of the data around a wavelength of 6400\,\AA\ within CD\,V. The log of all observations is given in the Appendix in Table~\ref{T1-App}.

Data reduction was done with the software package SDS4PEPSI (``Spectroscopic Data Systems for PEPSI'') based on Ilyin (\cite{4A}), and described in some detail in Strassmeier et al. (\cite{sun}). The specific steps of image processing include bias subtraction and variance estimation of the source images, super-master flat field correction for the CCD spatial noise, scattered light subtraction, definition of \'echelle orders, wavelength solution for the ThAr images, optimal extraction of image slicers and cosmic spikes elimination, normalization to the master flat field spectrum to remove CCD fringes and the blaze function, a global 2D fit to the continuum, and the rectification of all spectral orders into a 1D spectrum.

%------------------------F2
   \begin{figure*}
   {\bf a.}\\
   \includegraphics[width=\textwidth,clip]{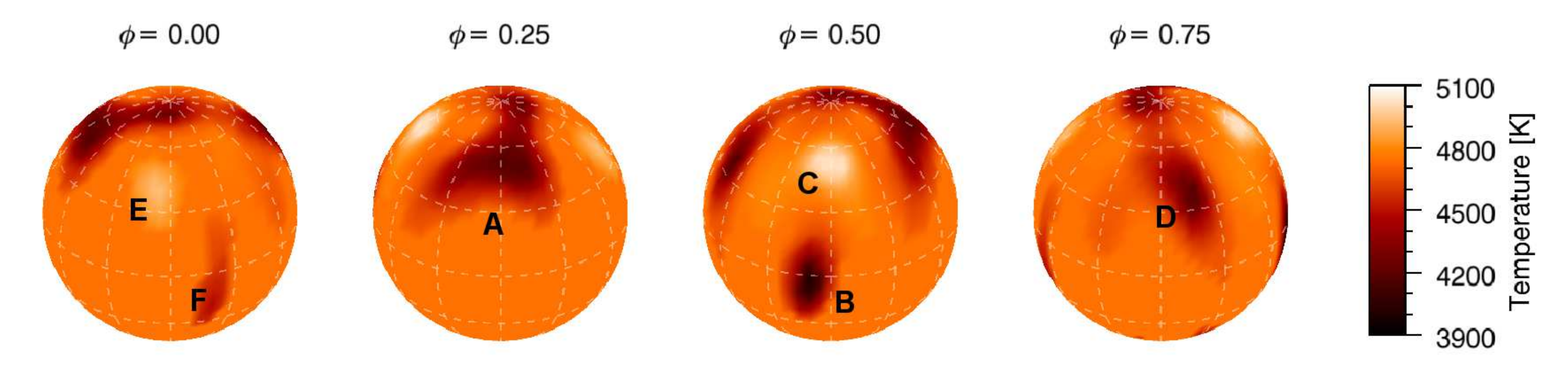}
   {\bf b.}\\
   \includegraphics[width=\textwidth,clip]{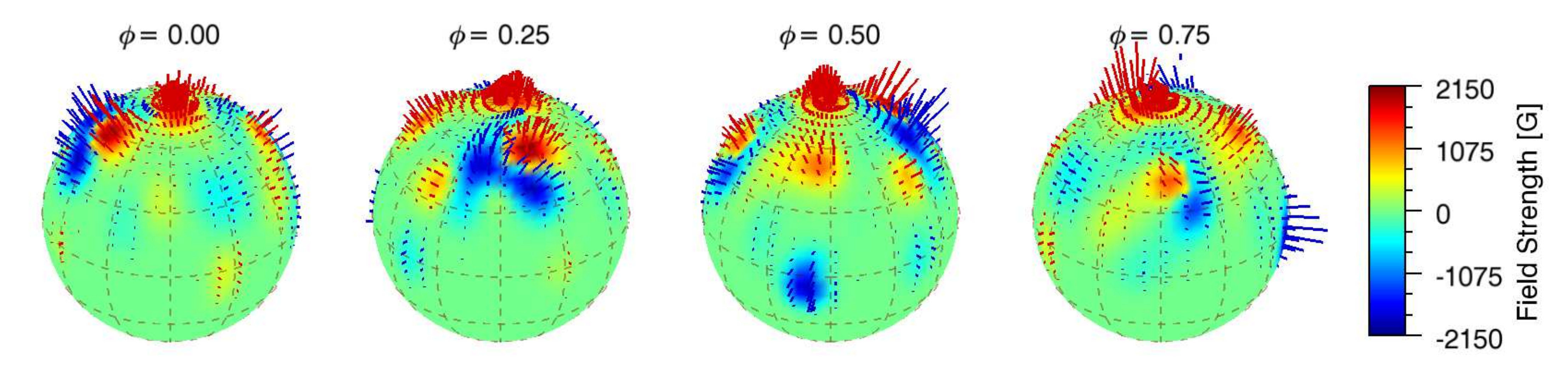}
   \caption{Doppler images of II~Peg.
     \emph{Panel a.} Temperature image.
     \emph{Panel b.} Magnetic-field image. $\phi$ is the rotational phase.
     Spot denominations in the top panel are: at phase $\phi$=0.25 the large cool feature is called spot~A; at phase $\phi$=0.5 the equatorial cool feature is called spot~B and the high-latitude warm feature spot~C; at phase $\phi$=0.75 the cool feature is called spot~D; and at phase $\phi$=0.00 the high-latitude warm feature is called spot~E and the equatorial cool feature called spot~F.}
         \label{F2}
   \end{figure*}

%------------------------F3
   \begin{figure*}
   \includegraphics[width=\textwidth,clip]{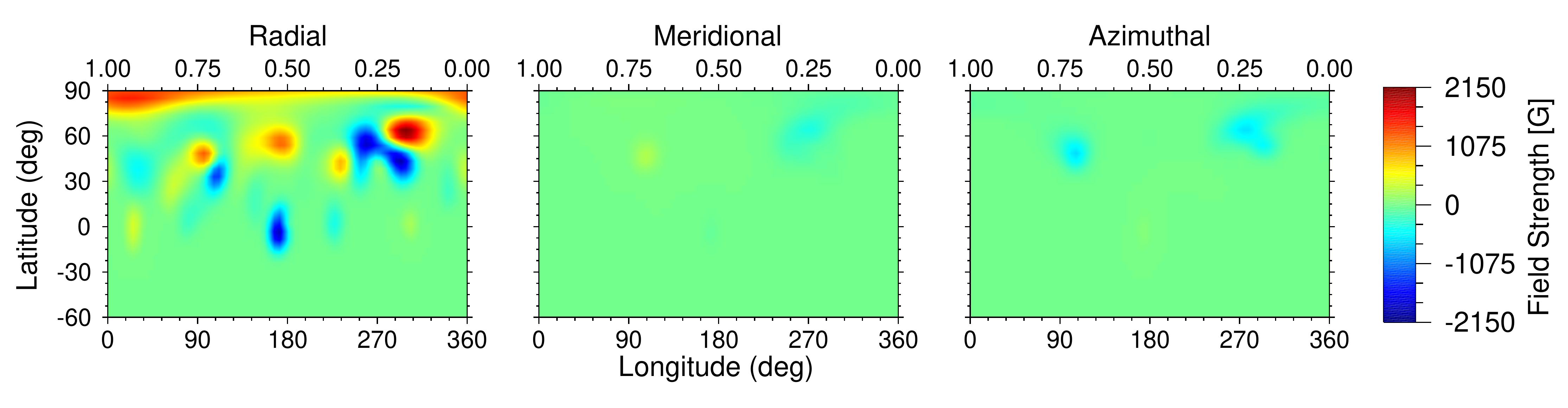}
   \caption{As in Fig.~\ref{F2}b but in Mercator-style projection and split into the three vector components. From left to right: radial-, meridional-, and azimuthal component. }
         \label{F3}
   \end{figure*}

%--------------------------------------------------------------------
\section{Zeeman Doppler imaging}\label{S3}

\subsection{Stellar parameters of II Peg}

Table~\ref{T1} summarizes the adopted stellar input parameters for II~Peg. We basically adopt the same parameters already used in the ZDI study by Ros\'en et al. (\cite{rosen15}) with some small improvements. The most important changes are a lower projected rotational velocity of 21.6\,\kms\ instead of 23\,\kms\ and a radial-tangential macroturbulence $\zeta$ of 3.6\,\kms\ instead of 4.0\,\kms\ as determined previously by Ottmann et al. (\cite{ott}) and Kochukhov et al. (\cite{koc:man}).

% ------------------------------ Table 1
\begin{table}
\caption{Astrophysical properties of II~Peg. } \label{T1}
\begin{tabular}{lll}
\hline \noalign{\smallskip}
Parameter                   & Value   & Based on  \\
\noalign{\smallskip}\hline \noalign{\smallskip}
Classification, MK          & K2\,IV  & various  \\
Effective temperature, K    & 4750 &  Ros\'en et al. (\cite{rosen15})  \\
Log gravity, cgs            & 3.5  & Ros\'en et al. (\cite{rosen15}) \\
$v\sin i$, \kms             & 21.6$\pm$0.5 & this paper \\
Microturbulence, \kms       & 2.0 & Ros\'en et al. (\cite{rosen15}) \\
Macroturbulence, \kms       & 3.6 & this paper \\
Rotation period, d          & 6.7242078 & =orbital period \\
Inclination, deg            & 60$\pm$10 & this paper \\
Metallicity, [Fe/H]$_\odot$ & --0.25  & Ros\'en et al. (\cite{rosen15}) \\
Distance, pc                & 39.36$\pm$0.06  & Gaia DR-2 \\
Radius, R$_\odot$           & 1.87$\pm$0.08 & from $\pi, T_{\rm eff}, V_{\rm max}$  \\
                            & 3.3$^{+1.0}_{-0.3}$ & from $v\sin i, i, P_{\rm rot}$ \\
\noalign{\smallskip}\hline
\end{tabular}
\tablefoot{$\pi$ parallax, $T_{\rm eff}$ effective temperature, $V_{\rm max}$ maximum brightness in Johnson $V$; $v\sin i$ projected rotational velocity, $i$ inclination of the rotation axis with respect to the sky plane, $P_{\rm rot}$ rotation period.}
\end{table}

The new \emph{Gaia} DR-2 parallax for II~Peg ($\pi$=25.4046$\pm$0.0393~mas; distance 39.36$\pm$0.06\,pc) converts to a stellar radius of just 1.87$\pm$0.08~R$_\odot$ using $T_{\rm eff}$=4750$\pm$100\,K and a maximum visual brightness $V_{\rm max}$=7\fm30, along with canonical solar values ($M_{\rm bol}$=4\fm83 and $T_{\rm eff}$=5770\,K) and the assumption of zero interstellar absorption. On the contrary, the minimum radius from the measured projected rotational velocity ($v\sin i$=21.6$\pm$0.5\,\kms) and the very precise rotation period of 6.724~d is 2.87$\pm$0.07~R$_\odot$. A likely inclination of the rotational axis of $60\degr\pm10\degr$ converts the minimum radius to a likely radius of 3.3$^{+1.0}_{-0.3}$~R$_\odot$. The two radii determinations are thus incompatible. Only if $v\sin i$ would be around 15~\kms\ or, on the contrary, $T_{\rm eff}$$\approx$3500\,K would the two radii agree, which we can both safely exclude from our high-resolution spectra. Only if a distance of $\approx$67~pc is assumed (or a visual magnitude of 6\fm0 instead of 7\fm3) then the radii from the two methods would be in agreement. This severe discrepancy already existed for the \emph{Hipparcos} parallax (23.62$\pm$0.89~mas) but appears now with higher weight due to the extremely precise \emph{Gaia} parallax. We have no readily explanation other than that the parallax measures are fooled by the changing brightness center due to II~Peg's huge spots. Because of this discrepancy we can not convert our relative spot areas to absolute units.

The rotational phases $\phi$ in Table~\ref{T1-App} were calculated from the orbital ephemeris given by Ros\'en et al. (\cite{rosen15})
\begin{equation} \label{eq1}
HJD = 2,448,942.428  + 6.7242078 \times\ E \ ,
\end{equation}
where the period is the orbital period and the zero point is a time of maximum (positive) radial velocity. Note that photometric period determinations (e.g., Rodon\'o et al. \cite{rodono}) always agreed with the orbital period within their error bars and with reasonable estimates for differential surface rotation.

\subsection{iMAP specifics}

All image reconstructions in this paper are done with the $i$MAP code (Carroll et al. \cite{carr07}, \cite{carr12}). $i$MAP does not use a spherical harmonics expansion for the magnetic field description but a physical 3D magnetic-field vector (radial, meridional, azimuthal) per surface pixel. We also use an iterative regularization where the step size and an appropriate stopping rule provides the regularization of the inverse problem. The Stokes I and Stokes~V inversions are done simultaneously.

Our present inversion technique is thus penalty free and is based on the Landweber iteration to minimize the sum of the squared errors (for more details see Carroll et al. \cite{carr12} and references therein). It can be written in a concise vector notation as
\begin{equation}\label{eq2}
\frac{1}{2} \| \vec{I}(\vec{x})  - \vec{O} \|^2 \rightarrow min \: ,
\end{equation}
where $\vec{I}$ is the synthetic model profile over all spectral lines, wavelengths, and rotational phases, and $\vec{O}$ is the corresponding observation. The vector $\vec{x}$ contains all our free parameters of the model, that is  the temperature and the magnetic-field vector for each surface element. For the stopping rule, we use the respective largest standard error of the reconstructed Stokes profiles. We note that the simultaneous inversion is dominated by the error in the Stokes~V profiles.

The code can either perform multi-line inversions for a large number of photospheric line profiles simultaneously or use a single average SVD-extracted line profile (again described in detail in Carroll et al.~\cite{carr12}). For the present application, we use the former for Stokes~I and the latter for Stokes~V. Both approaches apply our SVD technique. An eigenvalue decomposition of the signal covariance matrix, that is a SVD of the observation matrix, emphasizes the similarity of the individual Stokes profiles and allows one to identify the most coherent and systematic features. Incoherent features like noise and line blends etc. will be dispersed along many dimensions in the transformed eigenspace. For II~Peg, we created SVD-denoised profiles from 1811 spectral lines for the temperature inversion and SVD-averaged profiles for the magnetic-field inversion. We use the same 1811 spectral lines for the Doppler image as well as for the Zeeman-Doppler image. The difference is that the observation matrix for the inversion in Stokes~I still consist of the 1811 lines individually, each one denoised with the use of all lines, while in Stokes~V a single SVD profile (per phase) is used. The 1811 spectral lines were selected upon line depths larger than 10\% with respect to the local continuum and as blend free as possible. The average wavelength of the full line list is 5068\,\AA\ and the average Land\'e factor is 1.21. The individual Stokes-I profile has a S/N of around 1000 per pixel as observed, and listed in Table~\ref{T1-App}, but S/N$\approx$13\,000 per pixel per SVD-averaged profile in Stokes~V. A total of 14 rotational phases fairly equally distributed are available for the inversion.

For the local line-profile computation $i$MAP solves the radiative transfer with the help of an artificial neural network (Carroll et al.~\cite{carr08}). The atomic parameters for the line synthesis are taken from the Vienna Atomic Line Database (e.g., Ryabchikova et al.~\cite{vald}). These are used with a grid of Kurucz ATLAS-9 model atmospheres (Castelli \& Kurucz~\cite{atlas-9}) for local line profiles in 1D and in LTE. The grid covers temperatures between 3500\,K and 8000\,K in steps of 250\,K interpolated to the gravity, metallicity, and microturbulence values from Table~\ref{T1}.

The stellar surface is partitioned into 5\degr~$\times$~5\degr\ segments, resulting in 2592 surface pixels for the entire sphere. At the average resolving power of $\lambda/\Delta\lambda$=130\,000 (2.3~\kms\ at 6000\,\AA ), and an average full width of the lines at continuum of $2 \ (\lambda/c) \ v\sin i$ = 0.92~\AA , we have 20 resolution elements across the stellar disk of II~Peg. This is twice as many compared to previous maps based on, for example, CFHT and ESPaDOnS.

% ------------------------ Table 3
\begin{table}
  \caption{Spots on II~Peg in October 2017.}
\label{T3}
\centering
\begin{tabular}{llllll}
\hline\hline\noalign{\smallskip}
Spot  & Long & Lat & $\Delta T$ & $A_{\rm \Delta T}$ & $B_{\rm max}$ \\
ID    & (\degr )  & (\degr ) & (K) & (\% ) & (G) \\
\noalign{\smallskip}\hline\noalign{\smallskip}
A & 285 & +50 & +550 & 9.7 & --2000 \dots +1500\\
B & 170 & --5 & +780 & 2.8 & --2000 \\
C & 180 & +50 & --350& 1.9 & +1100 \\
D & 95  & +40 & +500 & 3.7 & --1100 \dots +1100\\
E & 350 & +35 & --150& 2.2 & +500 \\
F & 25  & --15& +150 & 2.2 & +600 \\
\noalign{\smallskip}\hline
\end{tabular}
\tablefoot{Longitudes and latitudes are given for the approximate spot centers. The spot temperature contrast, $\Delta T = T_{\rm phot}-T_{\rm spot}$, refers to the coolest (positive $\Delta T$) and warmest (negative $\Delta T$) part of the spot, while spot area $A$ is given for the entire size of the spot down to a threshold contrast of 50\,K in per cent of the visible hemisphere.}
\end{table}

\subsection{Results}\label{S33}

Our final DI and ZDI maps are shown in spherical projection in Fig.~\ref{F2} and in Mercator-style projection in Fig.~\ref{F3}. The line-profile fits are given in the Appendix in Fig.~\ref{F1-App}. The inversion was first done with Stokes~I only. Its solution was then used as the starting solution for the simultaneous Stokes-IV inversion. $i$MAP performed a total of about 3000 iterations for the final solution.

\subsubsection{Temperature map}

In terms of surface area the temperature map is dominated by a long and elongated cool spot group that extents via the polar regions across adjacent longitudinal hemispheres, but possibly bypasses the actual rotational pole itself. We call this feature spot~A when looked at phase 0.25 (identified in Table~\ref{T3}). Thus, no classical polar cap-like spot is seen during the epoch of our observation but rather an appendage toward the pole. The temperature of this spot complex appears different in different places. While the average temperature is only $\approx$450\,K cooler than the unspotted photosphere, it is $\approx$550\,K cooler in its central parts and fades off to $\approx$200\,K cooler on its equatorial rim. The connecting bridge around half of the visible pole is also only $\approx$250\,K cooler, but is still very well contrasted with respect to the pole itself. The area of the entire feature~A is 9.7\% of the visible hemisphere if we assume a likely temperature threshold of 50\,K with respect to the photospheric temperature. We note that spot areas in this paper are computed with the formula given in K\"unstler et al. (\cite{xxtri}) and based on a maximal contrast of $(\Delta T)_{\rm max} = (T_{\rm phot}-T_{\rm spot})_{\rm max} = 780$\,K in units of the visible hemisphere.

In terms of contrast the DI map is dominated by a pair of relatively round and compact-looking spots of which one is very cool ($\Delta T$$\approx$+780\,K) and the other very warm ($\Delta T$$\approx$--350\,K). The cool feature appears right on the stellar equator while the warm feature is at a latitude of +50\degr . Their respective central longitudes are only different by 10\degr\ in the sense that the warm feature is leading in the direction of stellar rotation. The relative areas are 2.8\% and 1.9\% for the cool and the warm spot, respectively. We call this pair spot~B (the cool one) and spot~C (the warm one). These two features will later be of central interest. Both spots appear isolated from the other spotted regions and are reconstructed with relatively well-defined edges compared to other features.

A fourth feature, spot~D, is located near a longitude of $\approx$95\degr\ (phase $\approx$0.8). It appears elongated and tilted with respect to a meridional line and could even be seen as an extension of spot~A via the pole into the adjacent opposite hemisphere. Its center near a latitude of +40\degr\ is reconstructed with a temperature contrast of $\approx$+500\,K and an area of 3.7\% if considered an isolated feature. This spot dominates the stellar disk when seen at phase 0.75 in Fig.~\ref{F2}a.

The remaining surface features are all comparably weak, yet still significant in terms of the achieved $\chi^2$ statistics. Note that the profile-fit level approaches the S/N of the data of $\approx$1000 per line to a very high degree (see Sect.~\ref{S41}). In an ideal world, this converts to an average temperature threshold of $\approx$10\,K (see Carroll et al. \cite{carr12}). Remaining conservative, we estimate our threshold to be closer to 30\,K, mostly limited by the imperfect phase sampling and the still limiting spatial resolution in spite of 14 phases with 20 resolving elements across the disk per profile. Nevertheless, two weak features shall be mentioned here: the warm spot~E with an area of 2.2\% and a contrast of $\Delta T$$\approx$--150\,K, and the cool suspiciously stretched spot~F, also  with an area of 2.2\%, but a maximum temperature contrast of $\Delta T$$\approx$+150\,K. It may be worth mentioning that the two warm features on the surface of II~Peg (spots C and E) are almost exactly 180\degr\ apart in longitude and approximately at the same latitude.

\subsubsection{Magnetic field map}

The visible polar area appears with a spatially offset positive field of strength up to 1200\,G, and is well correlated with a similarly offset cool region in the temperature map. The actual rotational pole is reconstructed with less than half of this field density ($\approx$500\,G; see Fig.~\ref{F3}), if at all, because the pole carries in principle no Doppler signal and positional errors are then largest. The field gradually converts to negative polarity at the adjacent lower latitudes at around $\approx$+60\degr\ where the polar field becomes comparably weak and never more than (--)100--200\,G. The field there is likely too weak and too complex that the Stokes~IV ZDI could capture its meridional or azimuthal components. The longitudinally adjacent side of the polar asymmetry (by nearly exactly 180\degr\ or at rotational phase $\approx$0.4) is where the high-latitude warm spot is reconstructed in the temperature map. This complexity prevents a clear statement of the pole's true magnetic morphology, that is whether it is also of bipolar or of singular polarity nature. Our reconstruction of the polar region gives an upper limit of the actual field density in a region of, say, 10\degr\ radius around the pole of around +1\,kG. We can not exclude the possibility of a mixed morphology with a weak negative polarity though.

In terms of field density, or strength, the map is dominated by the cool mixed-polarity spot~A and the cool negative-polarity spot~B. The warm positive-polarity spot~C with its peak value of +1100\,G is contrasted by the cool spots~A and B with peak values of $-2000$\,G. Most surprisingly, spots B \& C seem to form an opposite polarity pair at almost identical longitude but different latitude and appear to be the respective coolest and warmest features at that time on II~Peg. We are very confident of their reality because the two features are widely separated in latitude, were also reconstructed in the magnetic map, and the code has basically no unconstrained surface region due to incomplete phase coverage. If it were a typical artifact, for example caused by a faulty single line profile, the two hot-cool features would appear close together or even adjacent in latitude because the code tries to fit the faulty profile without bothering the other phases too much.

The cool and large spot~A is the most complex one and is reconstructed from Stokes~V with a detailed multi-polarity morphology. Four individual polarity regions are identified with peak field densities between +400\,G at the rotationally following side, --1100\,G in its central part, and +800\,G at the leading side. The coolest part of it is, as expected, also the strongest in terms of field density, and negative in polarity. The positive-field region of the leading side of spot~A is close to but not related with the second warm feature (spot~E, $\phi$$\approx$0.0) which is of same polarity but comparably weak ($\leq$200\,G). The ZDI map reveals a truly complex field morphology in this region (which is likely only the tip of an iceberg). Although all features shown in Fig.~\ref{F2}b are significant within the stated ZDI assumptions, the very weak magnetic feature are of course the most uncertain ones. Whether the one or the other of these weak features is an artifact can not be decided based on our data alone and we therefore add some caution to their interpretation. The cooler part of the large spot~D appears of solely negative polarity ($\approx$ --1000\,G) while the weaker and high-latitude part of it appears to be of positive polarity ($\approx$+1000\,G).

Overall, the high field densities strongly correlate with features in the temperature map. The coolest spots are typically the regions with highest field density. On the contrary, there are a few very weak (but still significant) radial field reconstructions even when no feature is seen in the temperature domain. For example, the cool spot~F in the temperature map is reconstructed with positive polarity of $\approx$300\,G and has a negative-polarity counterpart of same strength at higher latitudes, which itself has no counterpart in the temperature map. Another such feature is the appendage of spot~A toward the stellar equator at $\ell\approx 230\degr$ where a negative polarity of strength $\approx$300\,G is reconstructed on the equator but no feature is seen in the temperature image below a latitude of +30\degr. At this point we estimate our average local field-density threshold to be approximately $\pm$50\,G.

%--------------------------------------------------------------------
\section{Discussion and conclusions}\label{S4}

Here we discuss the robustness of our Stokes~IV maps and its magnetic-field morphology, and the possible explanations and consequences for the solar-stellar connection.

\subsection{Image robustness}\label{S41}

Figure~\ref{F1-App} shows the reconstructed set of Stokes V and Stokes I profiles compared with the SVD-averaged data for all 14 rotational phases. The standard errors are provided by the Bootstrap method with a re-sampling number of 1000. For each velocity bin, we obtained an estimate of the standard error which is condensed into a mean standard error averaged over the velocity domain. Its value is 9.2\,10$^{-4}$ for Stokes~I and 7.6\,10$^{-5}$ for Stokes~V, both values are basically identical with the respective inverse S/N due to the inversion stopping rule. To summarize, our arguments in favor of the robustness of the current ZDI map rests on the following items.

\emph{Initial data quality.} Even the stronger CP signal in spectral lines rarely exceeds 1\% of the continuum intensity in super active stars like II~Peg. Kochukhov et al.~(\cite{koc:man}) finds a Stokes~V signature  at the 2--3$\sigma$ confidence level in only a few of the strongest spectral lines of II~Peg in SOFIN data from the 2.4m NOT as well as in ESPaDOnS data from the 3.6m CFHT. Our PEPSI spectra in Fig.~\ref{F1} from the 11.8m LBT show Stokes~V signatures at the 10$\sigma$ level and were obtained with a twice as high a spectral resolution than previous maps of II~Peg. Given that the phase sampling has no significant voids, it resolves the stellar surfaces also twice as good. A simple test confirms this. We de-convolved the PEPSI data with an instrumental profile equal to the $R$=65\,000 spectral resolution as in previous ZDI work on II~Peg and then compared the map with that from the present data (Appendix Fig.~\ref{F2-App}). We see larger and less cool features in the lower resolution temperature map as well as less structured features of lower field strength in the magnetic map. Both differences, in area and in field strength, are approximately at the 10--20\% level. We recently found a similar impact with a comparable test for the $R=$250\,000 Stokes~I spectra of EK~Dra (J\"arvinen et al.~\cite{ekdra}).

\emph{Simultaneous T and $\vec B$ solution.}  In this paper, we had proceeded with an alternating minimization approach. That is, we started the magnetic inversion with an already pre-iterated temperature solution and then proceeded with the first magnetic inversion. After that, each iteration for the temperature was followed by an inversion for the magnetic field vector, which then used the information of the previous iteration of the temperature minimization. Therefore, the inversion starts with the step along the temperature gradient vector followed by a step along the magnetic gradient vector as laid out in Eq.~28 in Carroll et al. (\cite{carr12}). A total of 3000 such steps were performed. We found that the dependence of DI on the magnetic inversion is not strong while the dependence of ZDI on the temperature inversion is strong and, if completely neglected, leads to a non-trivial scaling of the Stokes~V profile amplitudes and widths and thus to a false ZDI map (Carroll et al. \cite{carr12}, Ros\'en \& Kochukhov~\cite{rosen12}). We also had done a test with different starting solutions for the alternate inversion. Instead of a pre-iterated initial temperature image, we used a blank surface with the effective temperature as the starting image for the Stokes~V inversion. The outcome was identical but more iterations were needed.

\emph{Penalty-free inversion.} Another important feature of our DI and ZDI is its iterative regularization. This allows us to make the inversion penalty free (for the expense of many iterations and thus CPU time). It means that we do not impose any constraints on the scale of the magnetic field, that is we do not indirectly prefer the large-scale surface structure over the small-scale structure by definition. The latter is the case for the previous ZDI inversions (Ros\'en et al. \cite{rosen15}) due to their adopted $l^2$ penalty function ($l$ is the angular harmonic expansion coefficient that describes the geometric scale of the magnetic field). In $i$MAP, only the surface gradient vector from one iteration to the next is used for the regularization.

\emph{Magnetic field density.} How does one convert the observed Stokes~V signal into a local magnetic field density (strength)? This is a non-trivial question and, so far, was implicitly answered through the assumption of the weak-field approximation (Stenflo \cite{stenflo94}, Carroll \& Strassmeier \cite{carr14}) from which we derive the effective and apparent longitudinal magnetic field. Due to the 3D vector description of $\vec B$ in $i$MAP, we do not need a conversion from harmonic expansion coefficients. We therefore believe that the radial-field densities reconstructed by $i$MAP from just Stokes~IV are closer to reality than previous Stokes~IV inversions and indirectly confirm the enhanced radial field strengths on II~Peg once Q\&U are added (Ros\'en et al. \cite{rosen15}). At this point we can not make a more quantitative statement regarding the even stronger meridional and azimuthal field enhancement from LP.

%------------------------F4
\begin{figure*}
   {\bf a.}\hspace{91mm}{\bf b.}\\
   \includegraphics[width=88mm,clip]{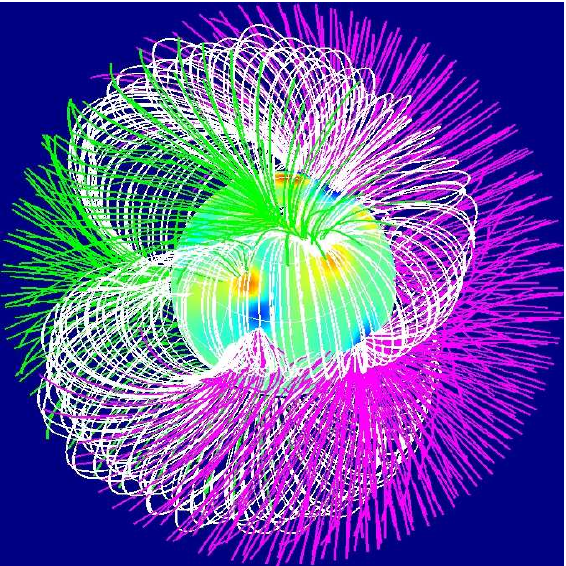}\hspace{4mm}
   \includegraphics[width=88mm,clip]{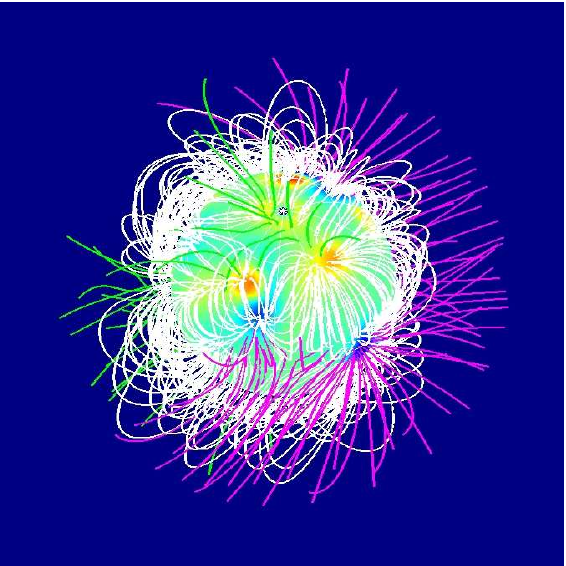}
   \caption{Magnetic-field extrapolation. Spherical projection for rotational phase 0.65. Open field lines are depicted in color (magenta negative polarity, red positive polarity), closed loops are in white. The ZDI map is color coded as in Fig.~\ref{F2}b. \emph{a.} Far-surface field with closed loops that reach a height of 2.2 stellar radii. \emph{b.} Near-surface field with closed loops that reach a height of 1.2 stellar radii. }
         \label{F4}
\end{figure*}

\subsection{Magnetic surface morphology}

We reconstruct surface features with a radial-field density of up to $-$2000\,G and +1500\,G, a meridional field density of up to $\pm$500\,G, and an azimuthal field density of up to $-$500\,G. This is to be compared to previous Stokes~IV results of II~Peg which reconstructed features of approximately $\pm$600\,G. However, Ros\'en et al. (\cite{rosen15}) found that the peak radial-field densities doubled when the linear component of the Stokes vector was included in their inversion, and even quadrupled for the other two components. Their IQUV-reconstructed peak local-field densities were therefore 1.3, 2.1 and 2.5~kG for the radial, meridional, and azimuthal field components, respectively. While the radial component of Ros\'en et al.'s and ours is comparable the meridional and azimuthal reconstructions are many times stronger than ours. The total magnetic energy is with 99\%\ almost exclusively in the radial component (with a magnetic flux of 86\%) while the meridional (0.6\%) and azimuthal (0.4\%) components are comparably weak. The field distribution is 53\%\ axisymmetric and 47\%\ non-axisymmetric.

Ros\'en et al. (\cite{rosen15}) argued that cross-talk between the radial and meridional field components can occur when only circular polarization is used in the magnetic inversion. This may be because Stokes~V is formally only sensitive to the line-of-sight component of the magnetic field. In Stokes~IV maps the radial component appears always to be the strongest of the three while it is the weakest in the IQUV inversion of Ros\'en et al. (\cite{rosen15}). We can not decide whether this is due to the missing linear polarization in our case, or related to the $l^2$ penalty function applied in the inversion by Ros\'en et al.. However, we have no analog from the Sun to compare with, or any readily physical effect, where the field density of a field parallel to the solar or stellar surface is higher than the radial-field density. We also note that the polar features in the magnetic maps by Ros\'en et al. changed polarity depending on whether Stokes IV or IQUV were used for the inversion.

We compute a 3D magnetic-field extrapolation based on our ZDI map as the inner boundary condition. An outer boundary condition is set as the potential field source surface (PFSS) and is defined through the Alfv\'en radius where the kinetic energy density equals the magnetic energy density. We set this source surface to three stellar radii. For the field extrapolation we use the potential field source surface model of Altschuler \& Newkirk (\cite{alt:new}) and Schatten et al. (\cite{schatt}). This model assumes that the field is potential and that the field lines are opened by the kinetic energy of the hot coronal gas at the source surface. For the actual calculation we use the SolarSoft PFSS package (Schrijver \& DeRosa \cite{trace}) that takes the radial component of the photospheric magnetic field from the ZDI map as input. Figure~\ref{F4} shows the extrapolations for the far-surface field ($<2.2$\,R$_\star$) and the near-surface field ($<1.2$\,R$_\star$) and for one and the same rotational phase. One can see that there are more closed field lines than open field lines, in particular the closer to the stellar surface one has chosen the outer boundary. Below $\approx$0.1 stellar radii the extrapolation becomes unreliable because it is based mostly on radial fields on the stellar surface.

\subsection{Spot origins}

Our most surprising result is the existence of a pair of well-defined compact warm and cool spots of high magnetic-field density but of opposite polarity; positive for the warm spot, negative for cool spot (spots C and B). A second warm spot, although less significant in terms of temperature contrast, appears also with positive polarity but consequently with somewhat lowered field density when compared to the dominating warm feature. The remaining cool spots on II~Peg appear with intermingled polarities.

Such a widely separated, bi-polar and cool/warm, photospheric feature has no direct solar analog. Cool solar spots are always formed due to a suppression of convection (Biermann \cite{bier}) while warm solar spots, called plages or faculae, are lower-density regions in the chromosphere usually physically located somewhat above the cool photospheric spots. In simple and well isolated active regions spots and plages are confined by the same magnetic field (e.g., Solanki~\cite{sol99}). However, solar surface reality is also that there are no active regions of apparently single polarity, only very young pores can appear with a single polarity (the other polarity still hidden in the surrounding convective layers). Moreover, sunspot polarities invert in the two hemispheres as well as in subsequent activity cycles (Hale's law). On II\,Peg, the bipolar feature (spots B\&C) spans a latitude range of 50\degr, thus almost an entire hemisphere on the visible stellar surface. Its cool spot is located almost exactly on the stellar equator while the warm spot is at +50\degr\ latitude. It resembles more a simplified solar coronal magnetic loop as seen in extreme UV images and movies (e.g., Schrijver \& DeRosa \cite{trace}) but enhanced in dimensions by a factor~10 or so.

%------------------------F5
\begin{figure}
   {\bf a.}\hspace{45mm}{\bf b.}\\
   \includegraphics[width=46mm]{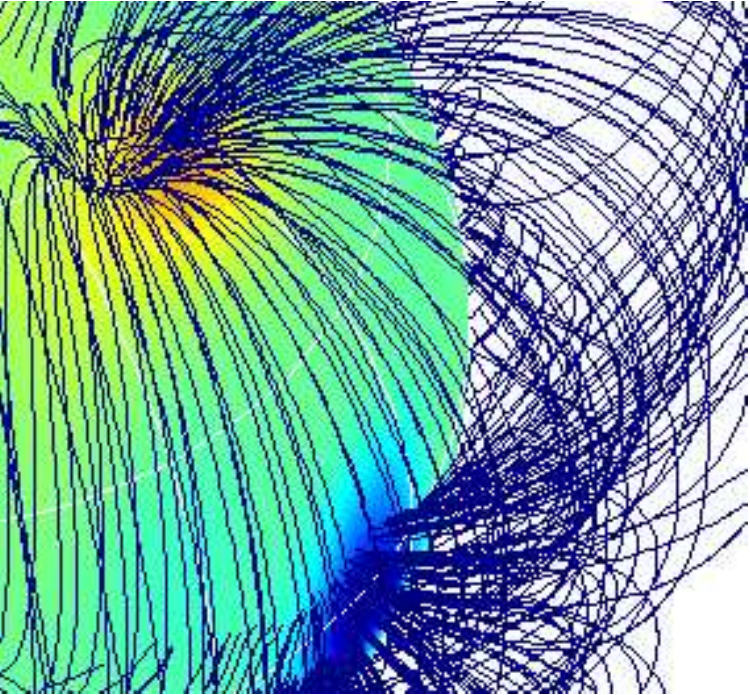}\hspace{1mm}
   \includegraphics[width=40mm]{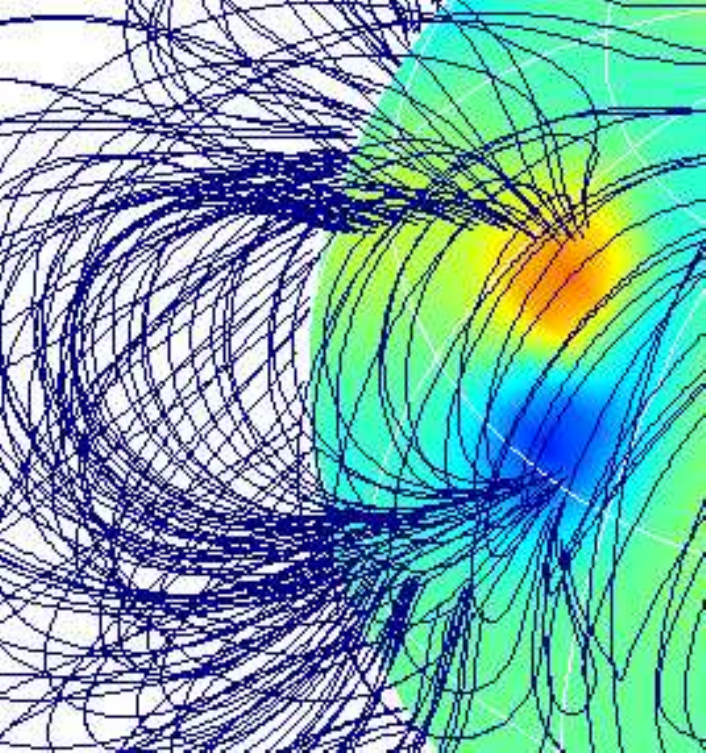}
   \caption{Loop structure for two bipolar spot pairs on II\,Peg. Shown are close-ups from the potential field extrapolation. \emph{a.} Spots B\&C (phase 0.65). \emph{b.} Spot D (phase 0.65)}
         \label{F5}
\end{figure}

Figure~\ref{F5}ab shows two excerpts from the magnetic-field extrapolation around the spot pair B\&C and spot~D, respectively. The B\&C pair consists of a warm (spot~C) and a cool spot (B) well separated, while spot~D appears as a single cool spot in the DI but is also reconstructed with dual polarity in the ZDI map just as the well-separated pair B\&C. Yet their connecting loop structure appears to be of similar height and geometry and is part of an even  larger loop pattern across the entire hemisphere (Fig.~\ref{F4}a).

We propose the following simple scenario to explain both the existence of warm and cool features and their bi-polarity. The field lines between spots B and C, or A and E, reconnect and form a closed loop. Within this loop electrons must be stripped from the negative polarity (on II~Peg always a cool spot) and be accelerated toward the positive polarity (on II~Peg always a warm spot). Eventually they shock near the loop end and form a hot plasma that heats the impact region beneath it. This region is then even visible in the optically-thin photospheric absorption lines as a warm spot in contrast to the cool background photosphere. This qualitative picture resembles very much the siphon flow model put forward to explain the Evershed effect in sunspots (e.g., Montesinos \& Thomas \cite{mon:tho}). However, Warnecke et al.~(\cite{war:che}) concluded that a one-dimensional coronal loop model, or a force-free extrapolation, cannot capture the complexity of the magnetic field in such a loop. Their simulations show that some field lines within the loop appear with twist angles of even 90\degr\ which would likely lower the electron acceleration rate between the polarities and possibly even prevent a classical shock front.

\begin{figure}
   \includegraphics[width=87mm]{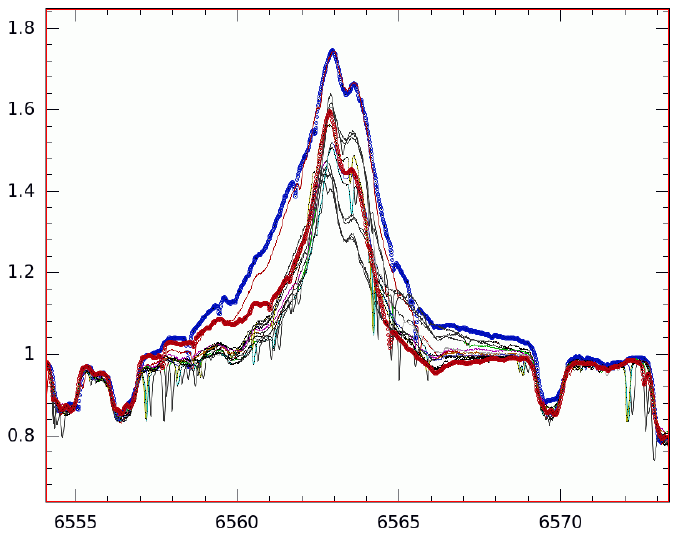}
   \caption{H$\alpha$ line-profiles changes during the time of the Doppler imaging (wavelengths in \AA). The blue-colored line profile is for phase 0.348 when the hot/cool spot pair B\&C is close to crossing the stellar central meridian. The red-colored line profile is the profile at the preceding phase 0.198 while the profile in-between these two profiles (thin line) is for phase 0.386. }
         \label{F6}
   \end{figure}

Some support for our loop scenario also comes from the Balmer-line behavior. Figure~\ref{F6} shows that when the warm feature (spot~C) is near crossing the central meridian the \Halpha\ line shows the strongest and widest residual emissions, this is also the case for the other Balmer lines from H$\beta$ through H$\epsilon$. Strongly variable Ca\,{\sc ii} H\&K, IRT, and He\,{\sc i} emission is evident in the spectra as well. The width of the H$\beta$ to H$\epsilon$ core emissions is between 2--3\,\AA\ (on average 150\,\kms) but 13~\AA\ for \Halpha. The Balmer \Halpha\ origin is clearly more complex, as initially already indicated by its non-Gaussian shape and large equivalent width  of --4.2\,\AA . Prior to the spot-C central-meridian passage its blue profile wing becomes more extended than the red wing while the central self reversal remains. All this indicates that most of the Balmer profile is formed locally far above the stellar photosphere, partly in the gas channeled along magnetic field lines behind the (assumed) siphon shock. Fairly constant He\,{\sc i} 5876-\AA\ emission during all phases suggests continuously existing high temperatures well above 10\,000\,K.

Besides the one particular pair of a warm and a cool spot, the majority of the cool regions on II~Peg appear containing both polarities, just like we would expect from a scaled-up solar active region. It is thus also likely that these features, in particular the multi-polar gigantic spot~A, are regions where the turbulent convective motion is suppressed and the flow of energy partly blocked or redirected. Therefore, the surface of II~Peg can host photospheric spots whose temperature originates from two rather different mechanisms at the same time, one related to a shock in a siphon-type flow and the other from the suppressed convection.

%------------------------F6

\subsection{Relation to the other binary component?}

The pair of warm and cold spots (spots B\&C) crosses the stellar central meridian very near phase 0.5, while the other warm feature (spot E) crosses the meridian very near 0.0. This alignment may be by chance but is within expected errors exactly 180\degr\ in separation, and at two distinct orbital locations. Because our phase zero point is a point of largest positive radial velocity, the two phases 0.0 and 0.5 refer to the two phases of quadrature. This is at least suggestive for a relation with the orbital position of the (unseen) secondary star. Because the orbit is circular there is no particularly exposed position in terms of tidal effects. The only plausible \emph{adhoc} explanation would be a magnetic link between the two stellar components (that would trigger flare activity). Clearly, this requires further evaluation in future work.

\section{Conclusions and summary}\label{S5}

Although observed many times in the past, II~Peg remains a stellar system with surprises. We found now good evidence that warm and cool features coexist on its surface. However, the warm feature(s) on II~Peg are likely not solar-analogs of plage or faculae-like features but are spatially well separated from the cool features. As such they are more like the foot points of magnetic-flux loops than a single up-scaled active region as for the Sun. Our ZDI map showed that warm features on II~Peg are of positive polarity, at least at the time of our observations, and that adjacent cool spots are of negative polarity. Other large cool areas were reconstructed with a mixed-polarity morphology. We interpret the loop connection to act like a siphon flow with a shock front that heats the local atmosphere near the loop end and within a certain height range.  When observed via optically-thin photospheric absorption lines, these regions then appears as warm spots. The simultaneous \Halpha\ line-profile variations are indicative of a mass flow between the two polarities, prematurely assuming a plasma flow from negative to positive polarity, at least for the spot pair B\&C. The field density, or strength, is dominated by far by the radial-field component reaching values of up to approximately $\pm$2\,kG. A comparably weak meridional and azimuthal component ($\approx$ $\pm$300--500\,G) is detected in our ZDI whenever opposite polarity regions are in close spatial vicinity. This is expected if the respective field lines indeed connect through loop-like structures as shown by our potential-field source-surface extrapolations. Future line-profile inversions will include also the LP components of the Stokes vector, as commonly used for high-resolution solar-surface inversions (e.g., Ruiz Cobo \& del Toro Iniesta \cite{ruiz}) and pioneered for stars like II~Peg by Ros\'en et al. (\cite{rosen15}). Finally, we note again the severe radius discrepancy from the parallax on the one hand and from the stellar rotation on the other hand.

\begin{acknowledgements}
LBT Corporation partners are the University of Arizona on behalf of the Arizona university system; Istituto Nazionale di Astrofisica, Italy; LBT Beteiligungsgesellschaft, Germany, representing the Max-Planck Society, the Leibniz-Institute for Astrophysics Potsdam (AIP), and Heidelberg University; the Ohio State University; and the Research Corporation, on behalf of the University of Notre Dame, University of Minnesota and University of Virginia. It is a pleasure to thank the German Federal Ministry (BMBF) for the year-long support for the construction of PEPSI through their Verbundforschung grants 05AL2BA1/3 and 05A08BAC as well as the State of Brandenburg for the continuing support of AIP and PEPSI (see https://pepsi.aip.de/). This work has made use of the VALD database, operated at Uppsala University, the Institute of Astronomy RAS in Moscow, and the University of Vienna.\\
This research has made use of NASA's Astrophysics Data System and of CDS's Simbad database which we both gracefully acknowledge.
\end{acknowledgements}

\appendix

\section{Additional plots and tables}

% ------------------------ T1-App
\begin{table*}
  \caption{Log of PEPSI observations.}\label{T1-App}
\centering
\begin{tabular}{lllllllll}
\noalign{\smallskip}\hline\hline\noalign{\smallskip}
Date  & HJD          & Total exp. & Stokes & \multicolumn{2}{c}{CD\,III} & \multicolumn{2}{c}{CD\,V} & $\phi$ \\
(UT)  & (2,458,000+) & (min)      &        & $\Delta\lambda$ (\AA )& S/N       & $\Delta\lambda$ (\AA )& S/N     & (Eq.~\ref{eq1}) \\
\noalign{\smallskip}\hline \noalign{\smallskip}
14/10/2017 &40.6288130& 20 & IV & 4800 - 5441 &   596  & 6278 - 7419 & 1202 & 0.052 \\
15/10/2017 &41.6140278& 20 & IV & 4800 - 5441 &   633  & 6278 - 7419 & 1230 & 0.198 \\
16/10/2017 &42.6196759& 20 & IV & 4800 - 5441 &   514  & 6278 - 7419 & 1025 & 0.348 \\
16/10/2017 &42.8734476& 25 & IV & 4800 - 5441 &   485  & 6278 - 7419 &  969 & 0.386 \\
10/10/2017 &36.6751209& 25 & IV & 4800 - 5441 &   397  & 6278 - 7419 &  791 & 0.464 \\
10/10/2017 &36.8003129& 20 & IV & 4800 - 5441 &   485  & 6278 - 7419 &  965 & 0.482 \\
10/10/2017 &36.9388681& 20 & IV & 4800 - 5441 &   361  & 6278 - 7419 &  816 & 0.503 \\
11/10/2017 &37.8191406& 15 & IV & 4800 - 5441 &   564  & 6278 - 7419 & 1064 & 0.634 \\
11/10/2017 &37.9113733& 20 & IV & 4800 - 5441 &   565  & 6278 - 7419 & 1160 & 0.648 \\
12/10/2017 &38.6196109& 20 & IV & 4800 - 5441 &   554  & 6278 - 7419 & 1128 & 0.753 \\
12/10/2017 &38.8294253& 20 & IV & 4800 - 5441 &   632  & 6278 - 7419 & 1278 & 0.784 \\
12/10/2017 &38.9365060& 20 & IV & 4800 - 5441 &   268  & 6278 - 7419 &  723 & 0.800 \\
13/10/2017 &39.6124016& 20 & IV & 4800 - 5441 &   608  & 6278 - 7419 & 1228 & 0.901 \\
13/10/2017 &39.8910861& 20 & IV & 4800 - 5441 &   470  & 6278 - 7419 & 1089 & 0.942 \\
\noalign{\smallskip}\hline
\end{tabular}

\tablefoot{The second column gives the heliocentric Julian date for the time of mid exposure, the third column is the total exposure time for both sub-exposures. S/N is per pixel and is the average from within the two wavelength regions $\Delta\lambda$. The last column is the rotational phase based on the ephemeris in Eq.~(\ref{eq1}).}
\end{table*}

%------------------------F1-App
   \begin{figure*}
   \centering
   \includegraphics[width=80mm]{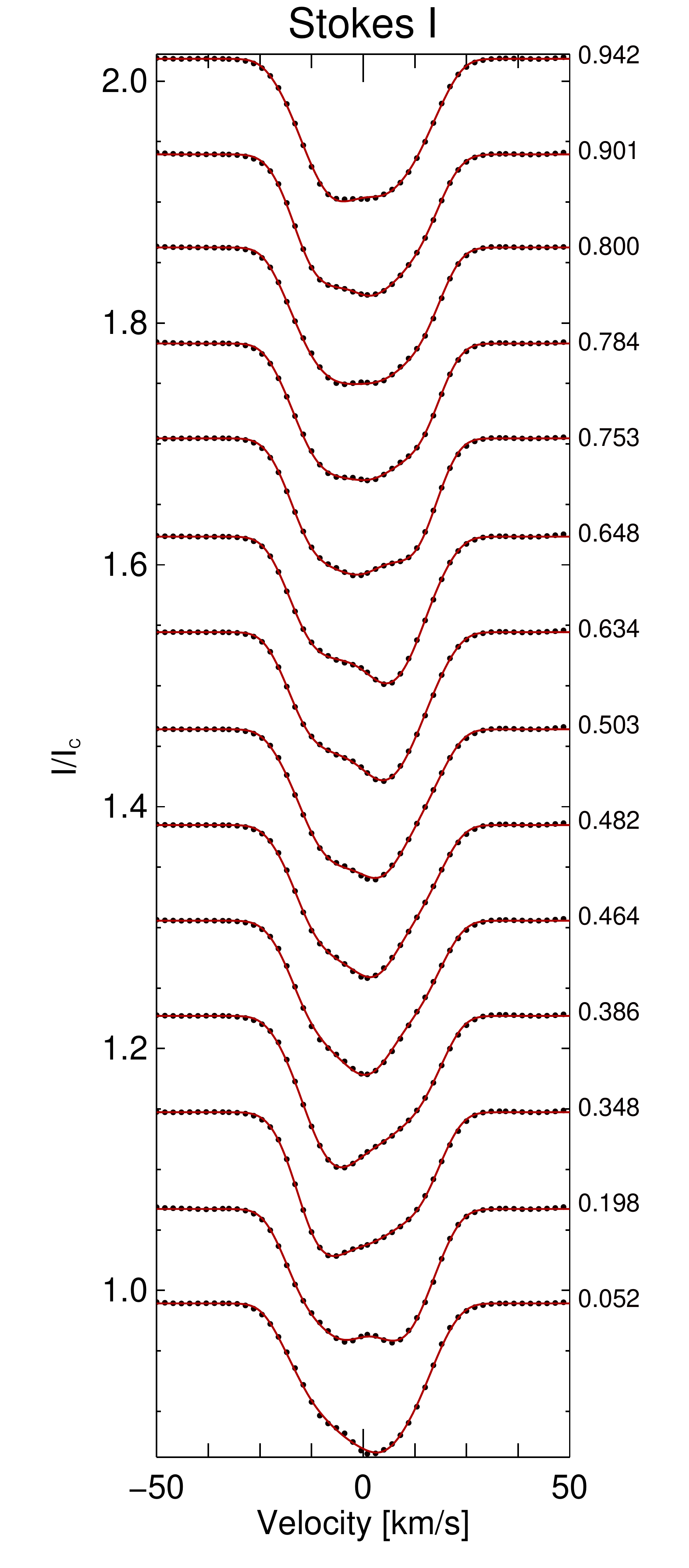}
   \includegraphics[width=80mm]{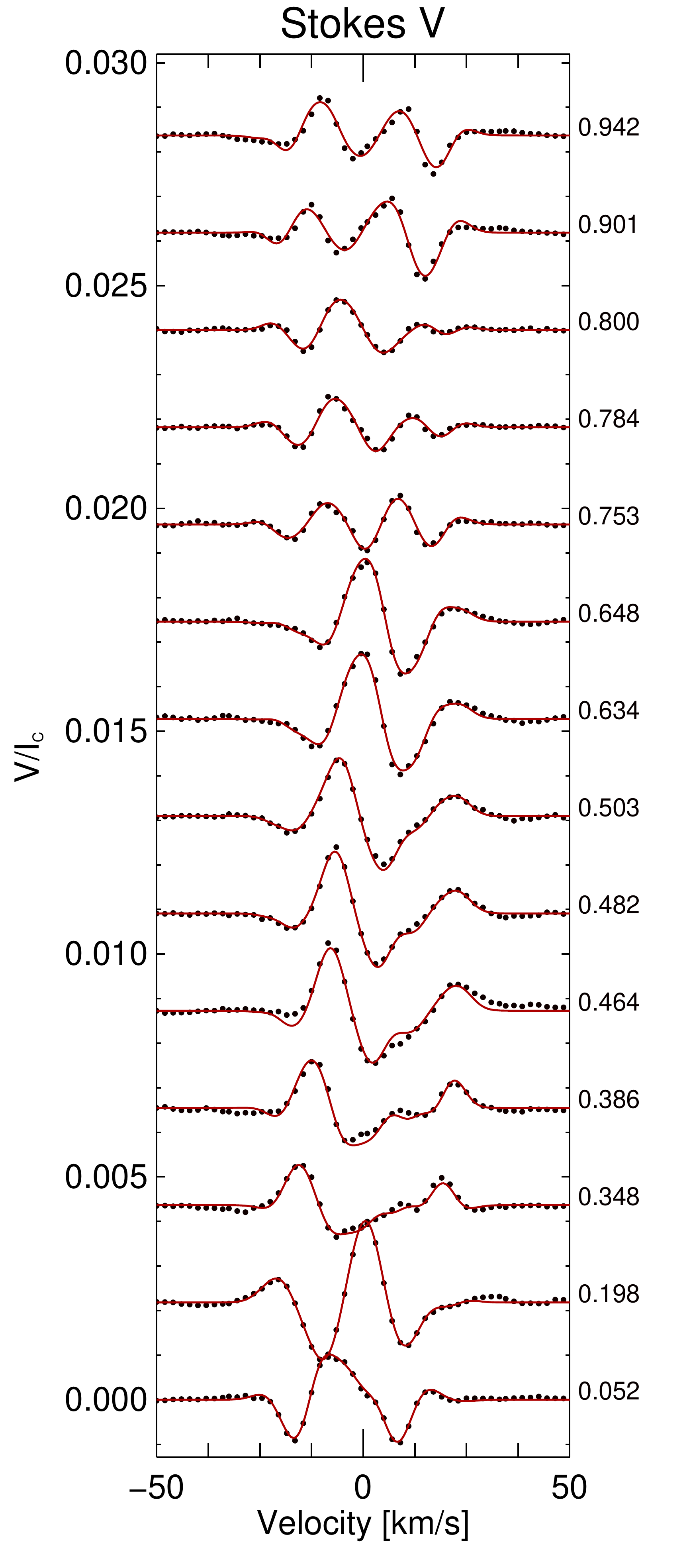}
   \caption{The observed (black dots) and inverted (red lines) line profiles
     for the temperature Doppler image (left) and the magnetic-field image (right). Profiles are
     labeled with their respective phases. Rotation advances from bottom to top.
              }
         \label{F1-App}
   \end{figure*}

%------------------------F2-App
   \begin{figure*}
   {\bf a. Original map with R=130\,000}\\
   \includegraphics[width=\textwidth, clip]{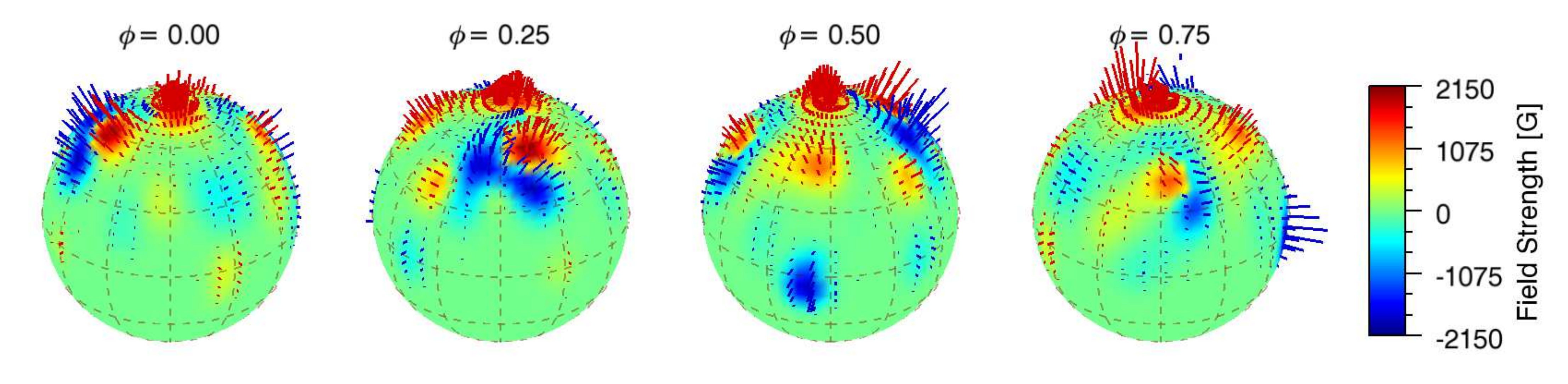}
   {\bf b. Map with R=65\,000}\\
   \includegraphics[width=\textwidth, clip]{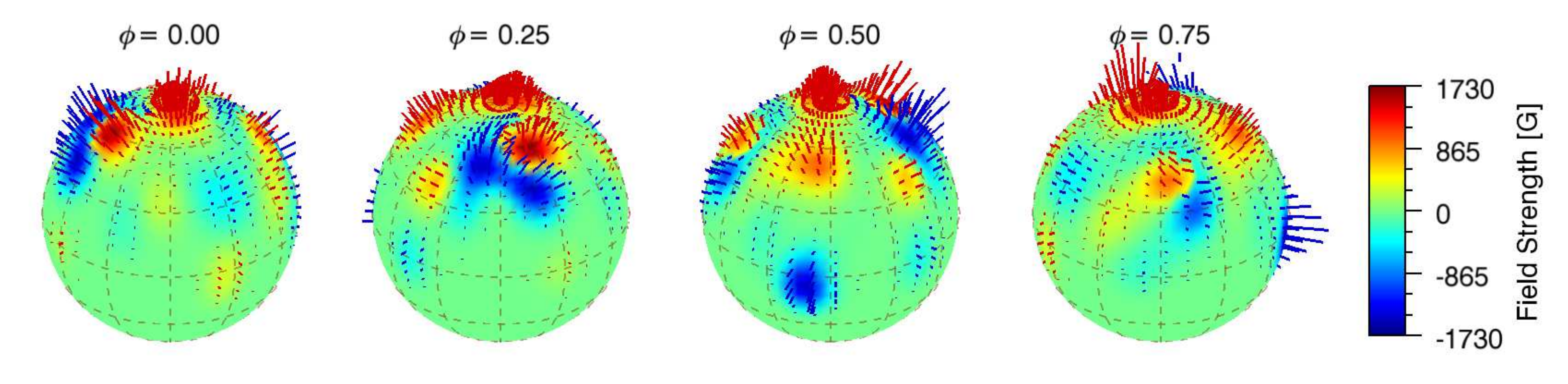}
   {\bf c. Difference map}\\
   \includegraphics[width=\textwidth, clip]{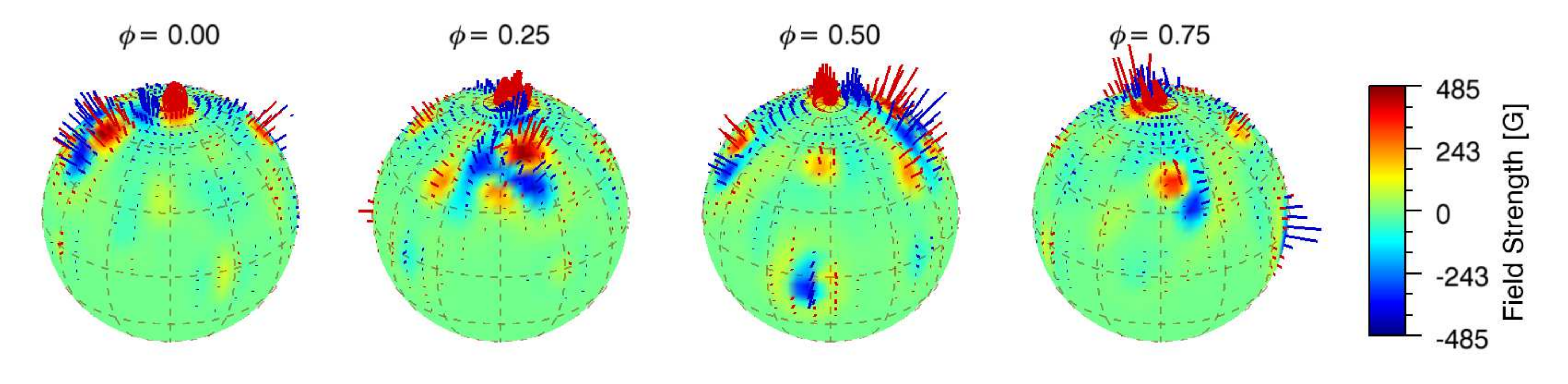}
   \caption{Zeeman Doppler images from data of different spectral resolution. Panel~a: based on the original data with a spectral resolution of $R$ = 130\,000 (as in Fig.~\ref{F2}b). Panel~b: based on the same data but downgraded to a spectral resolution of 65\,000. Panel~c: difference map $a-b$.
              }
         \label{F2-App}
   \end{figure*}


\begin{thebibliography}{}

\bibitem[1969]{alt:new}
Altschuler, M. D., \& Newkirk, G. Jr. 1969, SP, 9, 131

\bibitem[1998]{berd98}
Berdyugina, S., Berdyugin, A., Ilyin, I., \& Tuominen, I. 1998, A\&A, 340,
437

\bibitem[1999]{berd99}
Berdyugina, S., Berdyugin, A., Ilyin, I., \& Tuominen, I. 1999, A\&A, 350,
626

\bibitem[1938]{bier}
Biermann, L. 1938, AN, 264, 361

\bibitem[1987]{byrne}
Byrne, P. B., Doyle, J. G., Brown, A., Linsky, J. L., \& Rodono, M. 1987, A\&A, 180, 172

\bibitem[2007]{carr07}
Carroll, T. A., Kopf, M., Ilyin, I., \& Strassmeier, K. G. 2007, AN, 328, 1043

\bibitem[2008]{carr08}
Carroll, T. A., Kopf, M., \& Strassmeier, K. G. 2008, A\&A, 488, 781

\bibitem[2014]{carr14}
Carroll, T. A., \& Strassmeier, K. G. 2014, A\&A, 563, A56

\bibitem[2012]{carr12}
Carroll, T. A., Strassmeier, K. G., Rice, J. B., \& K\"unstler, A. 2012, A\&A, 548, A95

\bibitem[2004]{atlas-9}
Castelli, F., \& Kurucz, R. L. 2004, ArXiv Astrophysics e-prints
[astro-ph/0405087]

\bibitem[2009]{don:lan}
Donati, J.-F. \& Landstreet, J. 2009, ARA\&A, 47, 333

\bibitem[1992]{don92}
Donati, J.-F., Semel, M., \& Rees, D. E. 1992, A\&A, 265, 669

\bibitem[1991]{doyle}
Doyle, J. G., Kellett, B. J., Byrne, P. B. et al. 1991, MNRAS, 248, 503

\bibitem[2003]{gu}
Gu, S.-H., Tan, H.-S., Wang, X.-B., \& Shan, H.-G. 2003, A\&A, 405, 763

\bibitem[2012]{hack12}
Hackman, T., Mantere, M. J., Lindborg, M. et al. 2012, A\&A, 538, A126

\bibitem[1995]{hatz95}
Hatzes, A.P. 1995, in K. G. Strassmeier (ed.), Poster proceedings of IAU Symp. 176, University of Vienna, p.87

\bibitem[2000]{4A}
Ilyin, I. 2000, PhD Thesis, Univ. of Oulu, Finland

\bibitem[2012]{ilya12}
Ilyin, I. 2012, AN, 333, 213

\bibitem[2019]{ekdra}
J\"arvinen, S. P., Strassmeier, K. G., Carroll, T. A., Ilyin, I., \& Weber, M. 2019, A\&A, 620, A162

\bibitem[2013]{koc:man}
Kochukhov, O., Mantere, M. J., Hackman, T., \& Ilyin, I. 2013, A\&A, 550, A84

\bibitem[2015]{xxtri}
K\"unstler, A., Carroll, T. A., \& Strassmeier, K. G. 2015, A\&A, 578, A101

\bibitem[2011]{lind11}
Lindborg, M., Korpi, M. J., Hackman, T., et al. 2011, A\&A, 526, A44

\bibitem[1997]{mon:tho}
Montesinos, B., \& Thomas, J. H. 1997, Nature, 390, 485

\bibitem[1985]{mut:les}
Mutel, R. L., \& Lestrade, J. F. 1985, AJ, 90, 493

\bibitem[1998]{ott}
Ottmann, R., Pfeiffer, M. J., \& Gehren, T. 1998, A\&A, 338, 661

\bibitem[2012]{reiners}
Reiners, A. 2012, Living Reviews in Solar Physics, 9

\bibitem[2000]{rodono}
Rodon\'o, M., Messina, S., Lanza, A. F., Cutispoto, G., \& Teriaca L. 2000, A\&A, 358, 624

\bibitem[2012]{rosen12}
Ros\'en, L. \& Kochukhov, O. 2012, A\&A, 548, A8

\bibitem[2013]{rosen13}
Ros\'en, L., Kochukhov, O., \& Wade, G. A. 2013, MNRAS, 436, L10

\bibitem[2015]{rosen15}
Ros\'en, L., Kochukhov, O., \& Wade, G. A. 2015, ApJ, 805, 169

\bibitem[1977]{ruc}
Rucinski, S. M. 1977, PASP, 89, 280

\bibitem[1992]{ruiz}
Ruiz Cobo, B., \& del Toro Iniesta, J. C. 1992, ApJ, 398, 375

\bibitem[2015]{vald}
Ryabchikova, T., Piskunov, N., Kurucz, R. L., et al. 2015, Phys. Scr, 90, 054005

\bibitem[1969]{schatt}
Schatten, K., Wilcox, J., \& Ness, N. 1969, SP, 6, 442

\bibitem[2003]{trace}
Schrijver, C. J., \& DeRosa, M. L. 2003, SP, 212, 165

\bibitem[1984]{schr}
Schrijver, C. J., Mewe, R., \& Walter, F. M. 1984, A\&A, 138, 258

\bibitem[1989]{semel}
Semel, M. 1989, A\&A, 225, 456

\bibitem[1999]{sol99}
Solanki, S. K. 1999, in Butles, C. J. \& Doyle, J. G. (eds.), ASPC, 158, p.109

\bibitem[1989]{stenflo89}
Stenflo, J. O. 1989, A\&ARv, 1, 3

\bibitem[1994]{stenflo94}
Stenflo,  J.  O.  1994,  Solar  Magnetic  Fields:  Polarized  Radiation  Diagnostics, Astrophys. Space Sci. Lib.  (Dordrecht; Boston: Kluwer Academic Publishers), p.\,189

\bibitem[2009]{spots}
Strassmeier, K. G. 2009, A\&ARv, 17, 251

\bibitem[2015]{pepsi}
Strassmeier, K. G., Ilyin, I., J\"arvinen, A., et al. 2015, AN, 336, 324

\bibitem[2018]{spie-austin}
Strassmeier, K. G., Ilyin, I., Weber, M., et al. 2018, Proc. SPIE, 10702, 1070212

\bibitem[2018]{sun}
Strassmeier, K. G., Ilyin, I., \& Steffen, M. 2018, A\&A, 612, A44

\bibitem[1979]{vogt79}
Vogt, S. S. 1979, PASP, 91, 616

\bibitem[1980]{vogt80}
Vogt, S. S. 1980, ApJ, 240, 567

\bibitem[1992]{pfss}
Wang, Y.-M., \& Sheeley, N. R. 1992, ApJ, 392, 310

\bibitem[2017]{war:che}
Warnecke, J., Chen, F., Bingert, S., \& Peter, H. 2017, A\&A, 607, A53

\bibitem[2004]{web04}
Weber, M. 2004, PhD thesis, Univ. Potsdam

\bibitem[2013]{xi}
Xiang, Y., Gu, S.-H., Collier Cameron, A., \& Barnes, J. R. 2013, MNRAS, 438, 2307

\bibitem[2003]{zbor03}
Zboril, M. 2003, in N. Piskunov, W.W. Weiss, and D.F. Gray. (eds.), Poster proceedings of IAU Symp. 210, Uppsala University, p.\,D9

\end{thebibliography}
\end{document}